\documentclass[english,11pt,aps,prd,letterpaper,preprintnumbers,floatfix,nofootinbib,showpacs,superscriptaddress, notitlepage]{revtex4-1} \usepackage{amssymb}
\usepackage{slashed}
\usepackage{amsmath}
\usepackage{mathtools}
\usepackage{float}
\usepackage[utf8]{inputenc}
\usepackage{graphicx}
\usepackage{latexsym}
\usepackage{rotating}
\usepackage{epsfig}
\usepackage{xcolor}
\usepackage{natbib}
\usepackage[colorlinks=true,citecolor=darkred,urlcolor=darkred, pdfborder={0 0 0}]{hyperref}
\usepackage[normalem]{ulem}

\newcommand{\bea}{\begin{eqnarray}}
\newcommand{\eea}{\end{eqnarray}}

\definecolor{darkred}{rgb}{0.6,0,0}

\definecolor{linkcolor}{rgb}{0,0,0.5}

\newcommand{\realelpr}{\ensuremath{e_L^{80} p_R^{30}}}
\newcommand{\realerpl}{\ensuremath{e_R^{80} p_L^{30}}}
\newcommand{\pureelpr}{\ensuremath{e_L^{100} p_R^{100}}}
\newcommand{\pureerpl}{\ensuremath{e_R^{100} p_L^{100}}}

\begin{document}

\title{Probing heavy Majorana neutrino pair production at ILC
in a $U(1)_{\rm B-L}$ extension of the Standard Model}

\author{Jurina Nakajima}\email{jurina@post.kek.jp}
\affiliation{The Graduate University for Advanced Studies, SOKENDAI}
\affiliation{KEK, Tsukuba, Japan}

\author{Arindam Das}\email{arindamdas@oia.hokudai.ac.jp}
\affiliation{Institute for the Advancement of Higher Education, Hokkaido University, Sapporo 060-0817, Japan}
\affiliation{Department of Physics, Hokkaido University, Sapporo 060-0810, Japan}

\author{Keisuke Fuji}
\affiliation{KEK, Tsukuba, Japan}
\author{Daniel Jeans}
\affiliation{KEK, Tsukuba, Japan}
\affiliation{The Graduate University for Advanced Studies, SOKENDAI}
\author{Nobuchika Okada}
\affiliation{Department of Physics and Astronomy, University of Alabama, Tuscaloosa, USA}
\author{Satomi Okada}
\affiliation{Department of Physics and Astronomy, University of Alabama, Tuscaloosa, USA}
\author{Ryo Yonamine}
\affiliation{KEK, Tsukuba, Japan}

 \begin{abstract}

We consider a gauged B$-$L (Baryon number minus Lepton number) extension of the Standard Model (SM), 
which is anomaly free in the presence of three SM singlet Right Handed Neutrinos (RHNs). 
Associated with the $U(1)_{\rm B-L}$ gauge symmetry breaking, the RHNs acquire Majorana masses
and then with the electroweak symmetry breaking, tiny Majorana masses for the SM(-like) neutrinos are naturally generated
by the seesaw mechanism. 
As a result of the seesaw mechanism, the heavy mass eigenstates which are mainly composed of the SM-singlet RHNs 
obtain suppressed electroweak interactions through small mixings with the SM neutrinos.  
To investigate the seesaw mechanism, we study the pair production of heavy Majorana neutrinos
through the $U(1)_{\rm B-L}$ gauge boson $Z^\prime$ at the 250 GeV and 500 GeV International Linear Collider (ILC). 
Considering the current and prospective future bounds on the  B$-$L model parameters from the search
for a resonant $Z^\prime$ boson production at the Large Hadron Collider (LHC), 
we focus on a ``smoking-gun'' signature of the Majorana nature of the heavy neutrinos: 
a final state with a pair of same-sign, same-flavor leptons, small missing momentum, and  four hadronic jets.
We estimate the projected significance of the signature at the ILC. 

\end{abstract}

\maketitle

\vskip -0.3in 
\begin{center}
\rule[-0.2in]{\hsize}{0.01in}\\
 \vskip 0.1in 
 This is a preliminary study performed in the framework of the ILD concept group, \\
 submitted to the  Proceedings of the US Community Study\\
 on the Future of Particle Physics (Snowmass 2021)\\
 \rule{\hsize}{0.01in}\\
\end{center}


\def\thefootnote{\fnsymbol{footnote}}
\setcounter{footnote}{0}

\section{Introduction}

Although the SM postulated the neutrinos to be massless, experimental evidence of neutrino oscillations
indicates that SM(-like) neutrinos have tiny masses and flavor mixings \cite{Patrignani:2016xqp}. 
The \mbox{type-I} seesaw mechanism
\cite{Minkowski:1977sc,Mohapatra:1979ia,Schechter:1980gr,Yanagida:1979as,GellMann:1980vs,Glashow:1979nm,Mohapatra:1979ia} 
is probably the simplest idea to explain the origin of light neutrino mass generation, where the SM is extended with three SM gauge singlet Right Handed Neutrinos(RHNs). 
It has been known that the RHNs are naturally introduced in the minimal B$-$L (Baryon minus Lepton) model where the accidental global $U(1)_{\rm B-L}$ symmetry
in the SM is gauged. 
In addition to the SM particles, this model incorporates three SM singlet RHNs with $U(1)_{\rm B-L}$ charge $-1$ to cancel the gauge and
the mixed gauge-gravitational anomalies, an electric-charge neutral B$-$L gauge boson $(Z^\prime)$, and 
an SM singlet scalar with $U(1)_{\rm B-L}$ charge $+2$ called $U(1)_{\rm B-L}$ Higgs field.
The $U(1)_{\rm B-L}$ gauge symmetry is broken by a non-zero vacuum expectation value (VEV) of the B$-$L Higgs field, 
through which a mass for the $Z^\prime$ boson and Majorana masses for the RHNs are generated.  
With the Majorana RHNs, light neutrino masses are generated by the seesaw mechanism after electroweak symmetry breaking.
As a result of the seesaw mechanism, the heavy mass eigenstates which are mainly composed of the SM-singlet RHNs 
obtain suppressed electroweak interactions through small mixings with the SM neutrinos.

If the heavy Majorana neutrino mass eigenstates lie at the TeV scale or lower, they can be pair--produced in 
a process mediated by the $Z^\prime$ boson in the $s$-channel at high energy colliders such as the LHC and ILC.
Once pair-produced, each heavy Majorana neutrino decays to SM particles through its (suppressed) electroweak interactions
acquired through the seesaw mechanism. 
A leading decay mode of interest is to a $W$ boson accompanied by a charged (anti-)lepton, followed by two jets from hadronic $W$ decay.  
Due to the Majorana nature of the heavy neutrino, each heavy neutrino can produce either a charged lepton or anti-lepton. 
Therefore, heavy neutrino pair production can provide a very distinctive ``smoking-gun'' signature of its Majorana nature 
with a same-sign lepton pair, small missing energy, and four hadronic jets.

The gauge boson $Z^\prime$ has been sought at the LHC Run-2 and its production cross section is already severely constrained. 
The prospects for discovering the heavy Majorana neutrino in future LHC runs can be found in 
\cite{Kang:2015uoc,Cox:2017eme,Accomando:2017qcs,Das:2017flq,Das:2017deo,Jana:2018rdf,Das:2019fee,Chiang:2019ajm,Liu:2022kid}. 
A theoretical investigation for this ILC study can be found in~\cite{Das:2018tbd},
in which the authors point out the possibility that even after a null $Z^\prime$ boson search result at the High-Luminosity LHC, 
the 250 GeV ILC can search for heavy Majorana neutrino pair production to explore the origin of neutrino mass generation
through the seesaw mechanism.


\section{The B-L model}

Here we present a brief summary of the theoretical framework of the B$-$L model.  
New Yukawa couplings involving the RHNs are given by
\bea
-\mathcal{L}_{\rm int}\ \supset \ \sum_{i,j=1}^{3} Y_D^{ij} \overline{\ell_L^i} H N_R^j
  +\frac{1}{2} \sum_{i=1}^{3} Y_{N}^{j} \overline{(N_R^i)^{C}} \Phi N_R^i+ \rm{h. c.} \, , 
\label{U1xy}
\eea 
where $\ell_L^i$ and $N_R^i$ are the SM lepton doublet and RHN of the $i$-th generation ($i=1,2,3$), respectively,  
$H$ is the SM Higgs doublet, and $\Phi$ is the B$-$L Higgs field. 
Without loss of generality, we work in the flavor diagonal basis for the Majorana-type Yukawa couplings ($Y_{N}^{j}$) for RHNs.
We assume a suitable Higgs potential to yield VEVs for $\langle H \rangle =( \frac{v}{\sqrt{2}} \, \, 0)^T$ and 
$\langle \Phi \rangle =\frac{v_\Phi}{\sqrt{2}}$ to break the electroweak and $U(1)_{\rm B-L}$ symmetries, respectively.
Hence, the $Z^\prime$ boson mass, the neutrino Dirac mass term and the Majorana mass term for the RHNs are, respectively, 
generated as 
\bea
M_{Z^\prime} = 2 g_{\rm B-L} v_\Phi, \, \, M_D^{ij}=\frac{y_D^{ij}}{\sqrt{2}} v, \, \, M_N^i= \frac{Y_N^i}{\sqrt{2}} v_\Phi, 
\eea
where $g_{\rm B-L}$ is the B$-$L gauge coupling.
The mass matrix for the light SM(-like) neutrinos is generated to be $m_\nu \simeq m_D M^{-1} m_D^T$ by the seesaw mechanism. 
The heavy Majorana neutrino of the $j$-th generation acquire the gauge coupling with the $i$-th generation charged lepton/anti-lepton
and $W^+/W^-$ suppressed by a factor of  $m_{D}^{ij}/(M^j)$.  
A pair of ($i$-th generation) heavy Majorana neutrinos can be produced at the ILC with a center of mass energy $\sqrt{s}$,
whose production cross section is approximately given by  
\bea
\sigma(e^-e^+ \to Z^{\prime^\ast} \to N^i N^i)= \frac{s}{24 \pi} \Big(\frac{g_{\rm B-L}}{M_{Z^\prime}}\Big)^4 
\left(1-4\frac{M_{N^i}^2}{M_{Z^\prime}^2} \right)^\frac{3}{2}, 
\eea
where we assume $\sqrt{s} \ll M_{Z^\prime}$. 

\begin{figure}[t]
\centering
\includegraphics[width=0.75\textwidth,angle=0]{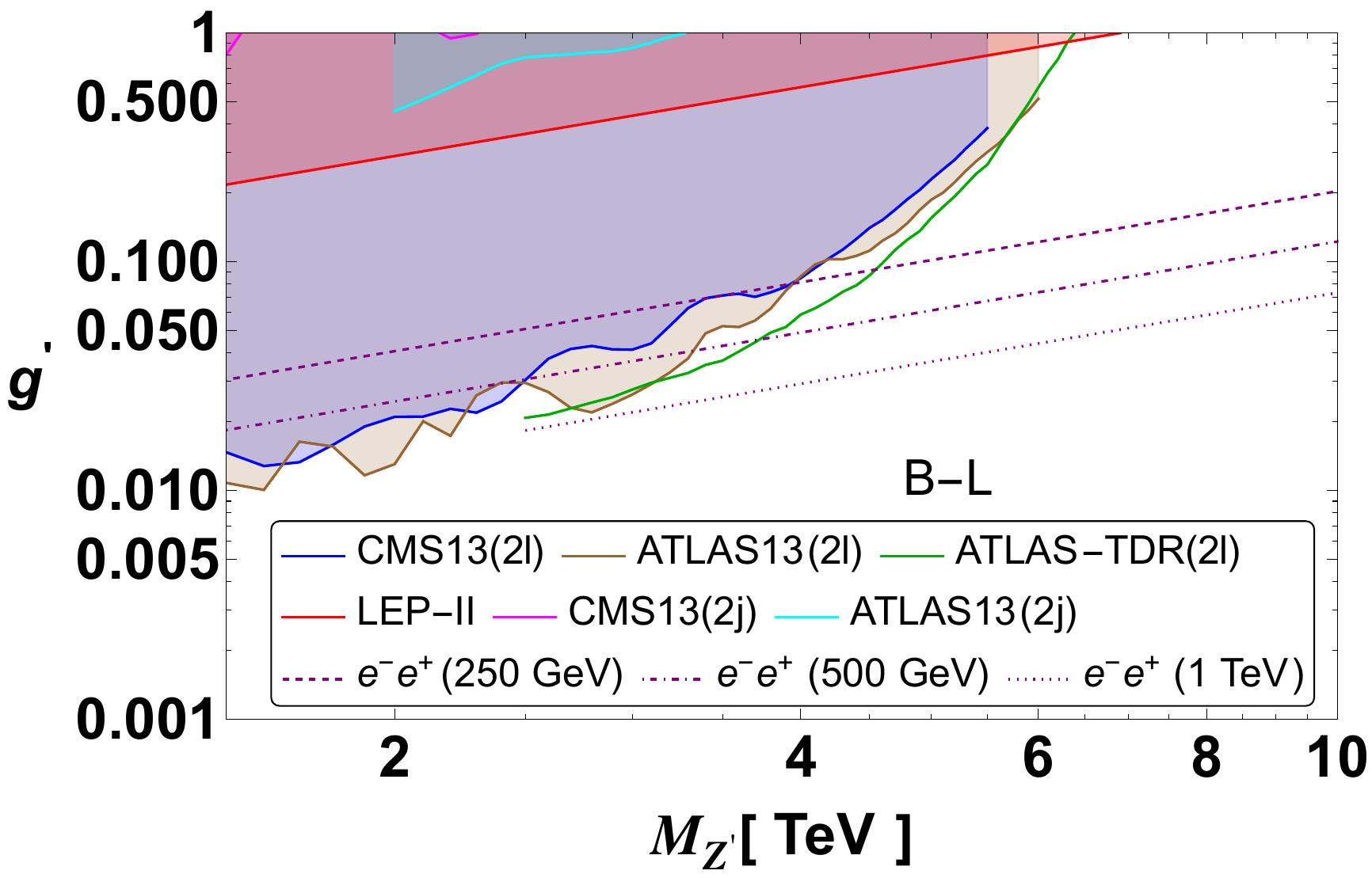}
\caption{
A summary of the upper bounds and prospective search reach on $g^\prime$ as a function of $M_{Z^\prime}$. 
}
\label{gp-MZp-1}
\end{figure}

The upper bound on the gauge coupling $g_{\rm B-L}$ as a function of $M_{Z^\prime}$ can be obtained from the LHC Run-2 results 
of the search for a narrow resonance with the dilepton final states by the ATLAS~\cite{ATLAS:2019erb} and 
the CMS~\cite{CMS:2021ctt} collaborations. 
The High-Luminosity LHC prospective reach is from ATLAS-TDR \cite{CERN-LHCC-2017-018}. 
The results of the search for a narrow resonance with the dijet final states by the ATLAS \cite{ATLAS:2019bov} 
and the CMS \cite{Sirunyan:2018xlo} collaborations are also used to derive the upper bound on $g_{\rm B-L}$. 
Considering the case when $M_Z^\prime > \sqrt{s}$ and using the limits on the effective scale for the left and right handed fermions from 
LEP-II \cite{Electroweak:2003ram} and ILC \cite{LCCPhysicsWorkingGroup:2019fvj} following \cite{Carena:2004xs,Okada:2016tci,Das:2021esm} 
we show the limits on the $g^\prime-M_{Z^\prime}$ plane from LEP-II and prospective ILC in Fig.~\ref{gp-MZp-1}.
The LEP-II results \cite{Schael:2013ita} also constrain $g_{\rm B-L}$ as a function of $M_{Z^\prime}$. 
These constraints are summarized in Fig.~\ref{gp-MZp-1}. 
For details on how to extract the upper bounds on $g_{\rm B-L}$ from various experimental results, 
see Ref.~\cite{Das:2021esm}, where the limits have been estimated using the generic $U(1)_X$ model file from FeynRules.
implementing the UFO model file in MadGraph \cite{Alwall:2011uj, Alwall:2014hca}. 
In Fig.~\ref{gp-MZp-1}, we also show the prospective search reach by the ILC with $\sqrt{s}=250$ GeV, $500$ GeV and $1$ TeV.

\section{Experimental study}

High energy electron-positron colliders such as the International Linear Collider ILC~\cite{Behnke:2013xla} have as their principal aim the 
study of the Higgs boson and electro-weak symmetry breaking.
Such a machine also provides a clean experimental environment and collision energies
never before seen in lepton collisions, which opens many opportunities to search for new
physics beyond the SM, either though precision measurements of known processes, or
the direct production of new particles.

The ILC is a proposed electron-positron collider which can provide several $ab^{-1}$
of data at centre-of-mass energies from 250 to 1000~GeV. The beams can be longitudinally
polarized to 80\% and 30\% for the electron and positron beams.
We refer to the case in which the electron (positron) beam is predominantly left(right)-handed as \realelpr,
and the opposite configuration as \realerpl.
In this study we consider the 500~GeV stage, ILC-500, and assume a run scenario in which
the total integrated luminosity at 500~GeV is $4 ab^{-1}$.

The International Large Detector ILD is a proposed detector concept for use at ILC~\cite{ILDConceptGroup:2020sfq}.
It consists of a high precision vertex detector, a hybrid tracking system based on a TPC with 
silicon layers, and highly granular calorimeters to enable optimal particle flow reconstruction. 
In this experimental study, we simulate RHN and SM background events in a detailed Geant-4 model of ILD.

\subsection{Signal and Background process}

We focus on the pair production of massive right-handed neutrinos, $e^{+}e^{-} \rightarrow NN$, at ILC-500. 
In the case of a RHN with mass greater than $m_H$, 
RHN can decay to $N\rightarrow l^\pm W^\mp, \nu Z$ and $\nu H$.
In this study we assume that the lightest RHN decays to the first lepton generation, 
and consider the decay $N\rightarrow e^\pm W^\mp$.


\begin{figure}[h]
    \centering
    \includegraphics{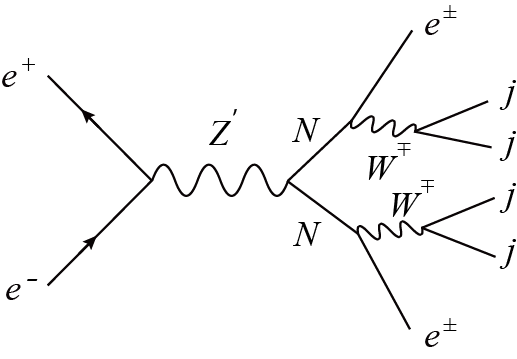} \hspace{5mm}
    \includegraphics{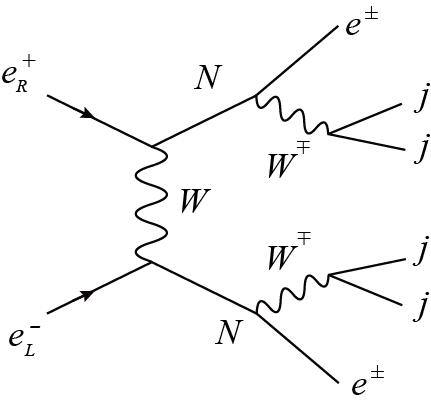}
    \caption{RHN pair production at tree level, with subsequent decay $N\rightarrow eW$, $W\rightarrow jj$.}
    \label{fig:Feynmandiagram_Signal}
\end{figure}
Feynman diagrams for $e^{+}e^{-} \rightarrow NN \rightarrow e^{\pm}e^{\pm}~+~4j$ are shown in Fig.~\ref{fig:Feynmandiagram_Signal}:
the process can occur either via s-channel $Z^\prime$ exchange or t-channel $W$ exchange. 
Since the RHN is a Majorana particle, in half of such decays, the electrons have the same sign. 
The cross-section for the s-channel process depends on the $Z^\prime$ mass $M_{Z^\prime}$, the coupling constant 
$g_{B-L}$ 
and the branching ratio $BR(N\rightarrow eW)$ that in turn depends on the mixing parameter $V_{eN}$
\begin{equation}
   \sigma(ee\rightarrow NN)\times (BR(N\rightarrow e^{\pm}W^{\mp}))^{2}  \propto \Big(\frac{g_{\rm B-L}}{M_{Z^\prime}}\Big)^4 \times BR(N\rightarrow e^{\pm}W^{\mp})^{2} , 
    \label{eq:exclusionlimit}
\end{equation}
while the t-channel process depends only on the $M_N$ and $V_{eN}$. 

\begin{figure}[ht!]
    \centering
    \includegraphics[scale=0.5]{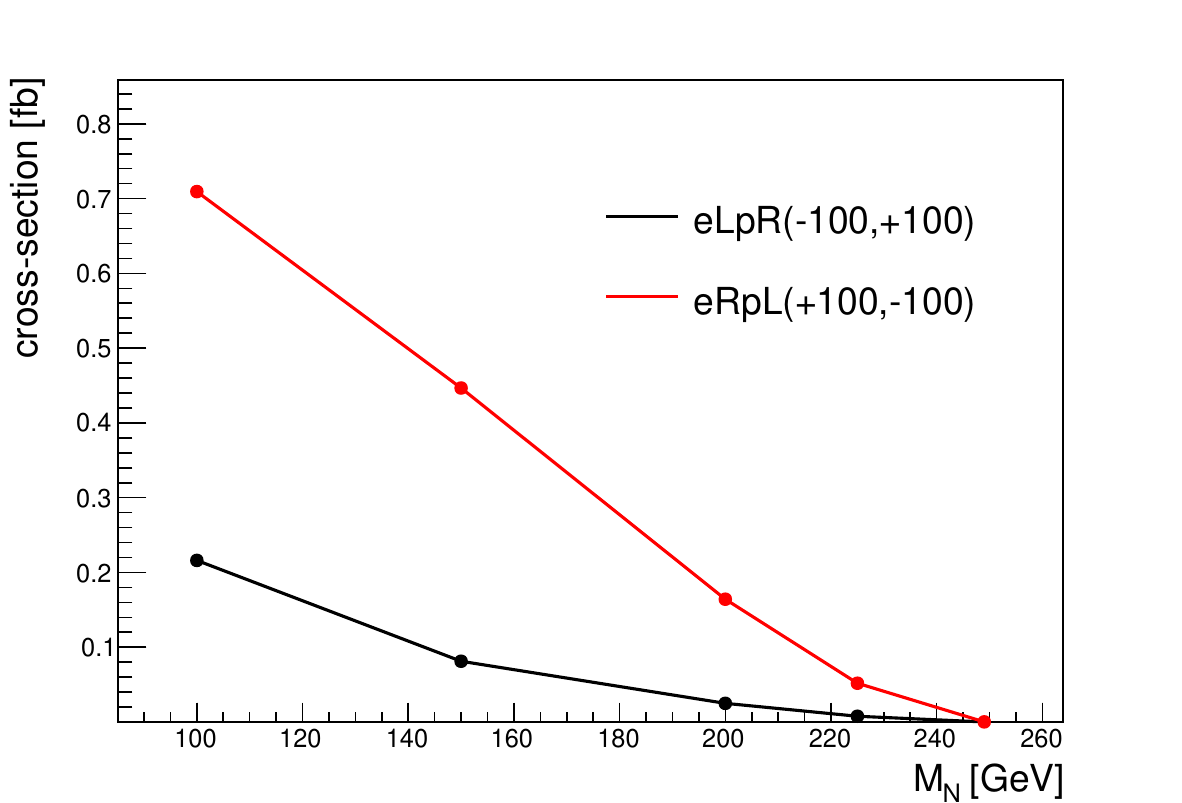}
    \caption{The cross-section $\sigma(ee\rightarrow NN)$ as a function of the RHN mass at the model point $g_{B-L} = 1, M_{Z^\prime} = 7$~TeV and $\mathrm{V}_{eN}$ = 0.03.}
    \label{fig:sigmavsmn}
\end{figure}

The cross-section as a function of RHN mass is shown in Fig.~\ref{fig:sigmavsmn} in the case of $g_{B-L} = 1$, $M_{Z^\prime} = 7~TeV$ and mixing parameter ${V}_{eN} = 0.03$,
for the 100\% polarized beam combinations \pureelpr\ and \pureerpl.
In the case of \pureerpl\ only the s-channel diagram contributes, since the right-handed electron does not couple to $W$. 
In the opposite polarization scenario both diagrams contribute, and the resulting cross-section is significantly suppressed due to destructive interference.
For this reason, in this study the results are mainly presented for the \realerpl\ polarization. 

Table~\ref{table:benchmark} shows the parameters of the four considered benchmark points. 
We study RHN masses ${M_{N}}$ in the range ${M_{N}} = 100 \sim 225~{GeV}$, allowing the RHNs to be pair-produced at ILC-500.
We assume $M_{Z^\prime} = 7$~TeV, which can be seen from Fig.~\ref{gp-MZp-1} is unconstrained by current measurements, 
and study various values of the $V_{eN}$ and $g_{B-L}$ model parameters.

\begin{table}[htb]
 \centering
\begin{tabular}{ccccccc}
\hline
${M_{N}}$ & ${M_{Z^{\prime}}}$ & $g_{B-L}^{\prime}$ & $|{V}_{eN}|$ & BR$(N\rightarrow eW)$ & $\sigma_{LR}$ & $\sigma_{RL}$ \\
$[\mathrm{GeV}]$   & $[\mathrm{GeV}]$  &  &      & & \multicolumn{2}{c}{[fb]}\\ 
\hline
100     & 7                & 1    & 0.03 & 0.44 & 0.55  & 0.71    \\
150     & 7                & 1    & 0.03 & 0.33 & 0.36  & 0.45    \\
200     & 7                & 1    & 0.03 & 0.30 & 0.14  & 0.16    \\
225     & 7                & 1    & 0.03 & 0.29 & 0.046 & 0.0052  \\ 
\hline
\end{tabular}
 \caption{Benchmark points considered in this analysis. Cross-sections assume 100\% beam polarization and take account of ISR and beamstrahlung.}
 \label{table:benchmark}
\end{table}


The final state in which we search for the signal has no irreducible backgrounds from Standard Model processes.
Possible sources of background occur when a lepton originating in hadron decay (particularly from heavy flavor hadrons)
is identified as an isolated lepton, and paired with a charged lepton from e.g. $W$ decay.
The potential for charge misidentification of the considered electrons was found to be completely negligible
in the full simulation of the ILD.
To estimate the SM backgrounds, 4-- and 6--fermion final states with at least one electron and two quarks are considered.

\begin{figure}[htb]
\begin{center}
  \includegraphics[width=0.35\textwidth]{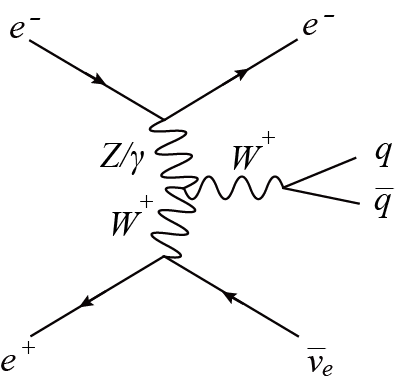}
  \includegraphics[width=0.35\textwidth]{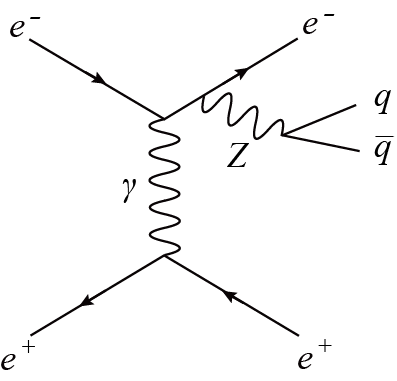} \\
  \includegraphics[width=0.45\textwidth]{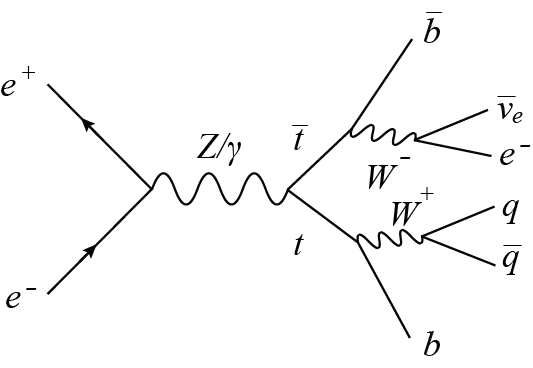}
  \caption{Examples of Feynman diagrams contributing to the main backgrounds}
  \label{bg_diagram}
\end{center}
\end{figure}

\begin{itemize}
    \item 4--fermion: $e^{+}e^{-}\rightarrow e\nu q\bar{q}$ and $e^{+}e^{-}\rightarrow e^{+}e^{-}q\bar{q}$
    \item 6--fermion: $e^{+}e^{-}\rightarrow e^{+}e^{-}q\bar{q}q\bar{q}$
    \item $e^{+}e^{-}\rightarrow t\bar{t}$
\end{itemize}
Feynman diagrams corresponding to some of these processes are shown in Fig.~\ref{bg_diagram}.
We expect these processes to provide the majority of backgrounds in this analysis, however
in a later update of this note additional processes which could potentially provide additional background contributions 
will be added.

Table~\ref{table:bg_sigma} shows the cross-sections for backgrounds. 
The cross-section of 4--fermion processes are significantly enhanced in the \pureelpr\ polarisation. 
\begin{table}[htb]
 \centering
\begin{tabular}{c|cc|cccc}
\hline
 Beam pol.         &  \multicolumn{2}{c|}{4-fermion} & \multicolumn{4}{c}{6-fermion} \\
 ($100\%$)  &  $e\nu qq$          & $eeqq$             & $eeqqqq$ & $t\bar{t} (2e)$ & $t\bar{t} (1e)$ & $t\bar{t}(0e)$  \\
\hline
$e_Lp_R$               & 7810 & 1960 & 16.63  & 20.1 & 313.2 & 1200 \\
$e_Rp_L$               & 22.8 & 1730 & 3.74   & 7.56 & 119   & 461  \\
$e_Lp_L$               & 753  & 1780 & 6.69   & 0.11 & 0.74  & 0    \\
$e_Rp_R$               & 750  & 1780 & 6.66   & 0.11 & 0.73  & 0    \\
\hline
\end{tabular}
 \caption{The cross-sections [in fb] for the SM backgrounds considered in this analysis. Beam polarization of 100\% is assumed.
The $t\bar{t}$ process is split into events with 0, 1, and 2 electrons.
}
 \label{table:bg_sigma}
\end{table}

\subsection{Simulation Setup}

Signal events were generated using the WHIZARD event generator 2.8.5~\cite{2011_Whizard}. 
The general model file is written in FeynRules and the UFO model of a general U(1) extension of the SM is written in \cite{Das:2022rbl}.
The B-L scenario can be obtained using $x_{H}=0$ and $x_{\Phi}=1$. 
The UFO file of the \href{https://feynrules.irmp.ucl.ac.be/wiki/GeneralU1\#no1/}{general U(1) model} file can be found from the FeynRules database.

We set very large masses for two of the RHNs and set the RHN mixing parameter to $\mu$ and $\tau$ leptons to zero. 
Samples of background events, prepared by the ILD software group, using the WHIZARD 1.9.5. 
Signal and background events include the effect of Initial State Radiation (ISR), 
as well as of beamstrahlung simulated by GuineaPig~\cite{Schulte:1999tx} based on ILC accelerator parameters~\cite{adolphsen2013international}.
WHIZARD generates the hard event, and the subsequent parton shower and hadronisation was simulated by Pythia v6.24~\cite{Sjostrand:2006za}.

DDsim~\cite{sailer2017dd4hep}, a software toolkit based on Geant4~\cite{GEANT4:2002zbu}, was used to simulate the interaction of final state particles in a detailed model of ILD. 
Reconstruction and analysis was performed in the Marlin analysis framework~\cite{Gaede:2006pj}, 
including track finding and fitting, and particle flow analysis using PandoraPFA~\cite{Thomson:2009rp}.
The IsolatedLeptonTagging processor was used to identify isolated electrons and muons, and particles other than isolated leptons and photons are clustered into jets using the Durham algorithm~\cite{Catani1991432}.

\subsection{Cut Based Analysis}

In this section, we present the event selection cuts designed to isolate the signal from SM backgrounds.
The most useful signature of the signal process is the presence, in half of the events, of a pair of same--sign isolated electrons, most usually together with 4 hadronic jets from $W$ decay. 
We impose the following selection cuts:
\begin{enumerate}
     \item Exactly two same-sign isolated electrons, and no isolated muons or photons.
     \item The energy of both isolated electrons should be $<$ 200 $\mathrm{GeV}$
     \item Both isolated electron polar angles $|\cos{\theta_{isoe}}| < $ 0.95.\\
       The dominant backgrounds involve t-channel processes, which tend to have forward-going electrons.
     \item IsolatedLeptonTagging parameter of both electrons $>$ 0.9.\\
        The IsolatedLeptonTagging parameter is the output of a Multivariate Data Analysis trained to identify isolated electrons. 
        Same sign background events almost always have one true and one mis-identified isolated electron. This cut reduces events with mis-identified electrons.
     \item Jet clustering parameter with Durham $\log_{10}{(y_{12})}$ $> -1$.\\
     To reduce events with less than 4 jets.
     \item Magnitude of the missing momentum $\mathrm{P}_{miss}$ $<$ 100 $\mathrm{[GeV]}$, and ($\mathrm{P}_{miss}$ $<$ 40 $\mathrm{[GeV]}$ or $|\cos{\theta_{\mathrm{P}_{miss}}}| > $ 0.95)\\
     Missing momentum in the signal is either small, or due to unseen ISR in the very forward region. 
     This cut removes events with leptonic $W$ boson decays.
\end{enumerate}

\begin{figure}[htb]
 \centering
    \includegraphics[width=0.48\textwidth]{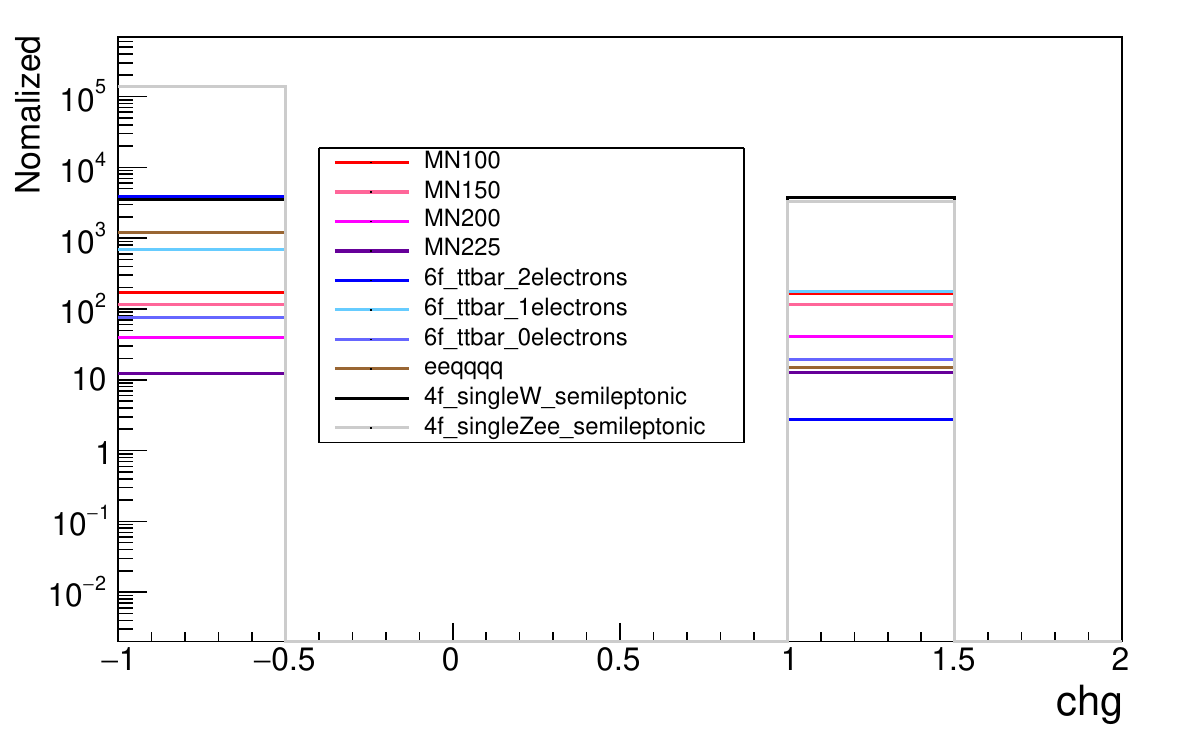}
    \includegraphics[width=0.48\textwidth]{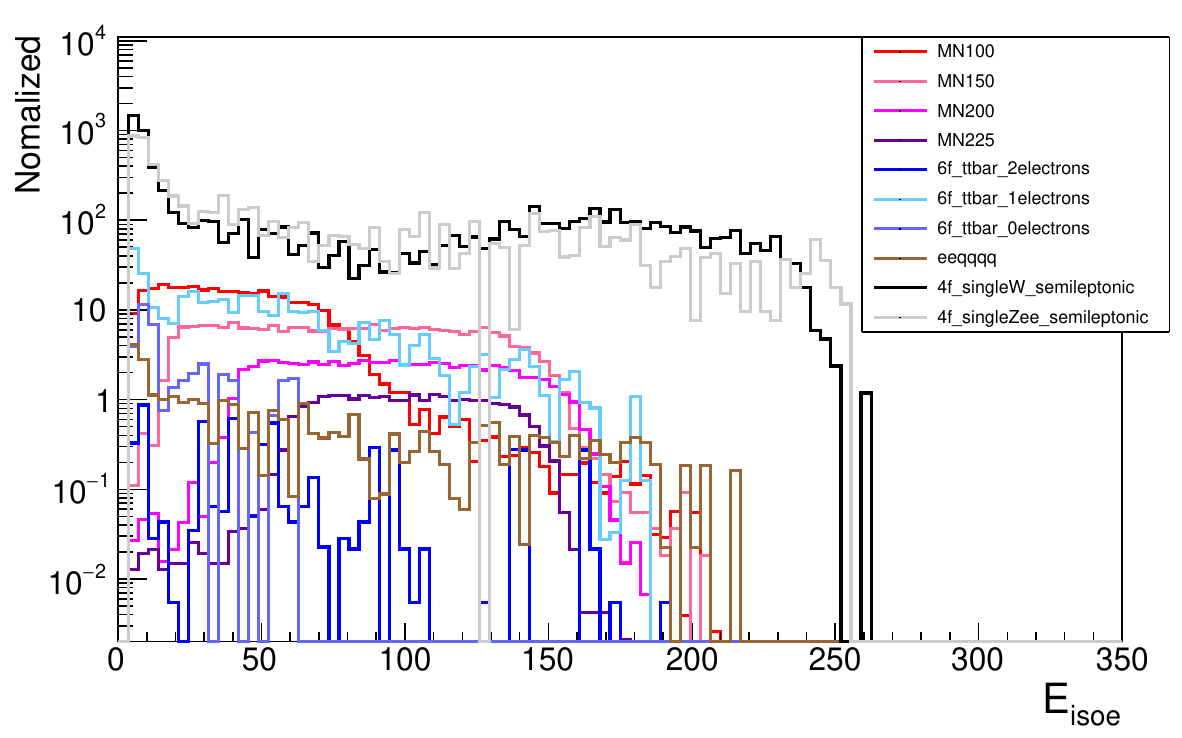} \\
\vspace{-55mm}
\hspace{30mm}{\small    ILD preliminary \hspace{0.3\textwidth}  ILD preliminary }\\
\vspace{55mm}
    \includegraphics[width=0.48\textwidth]{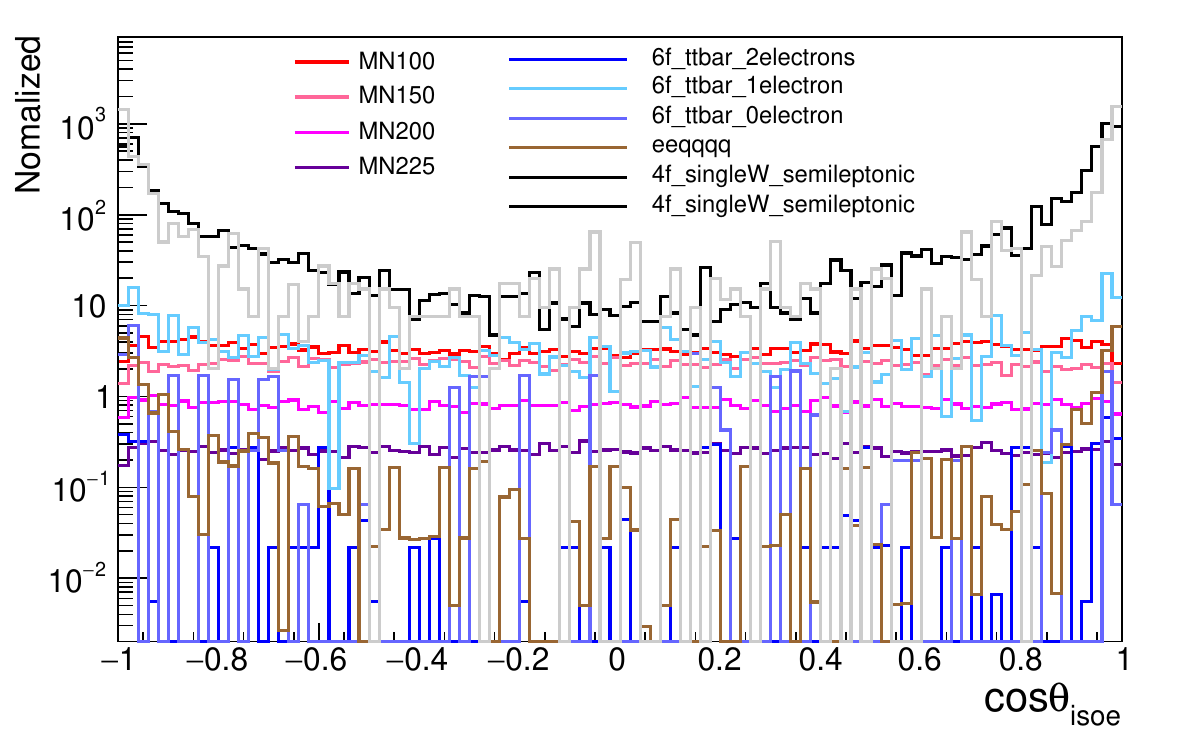}
    \includegraphics[width=0.48\textwidth]{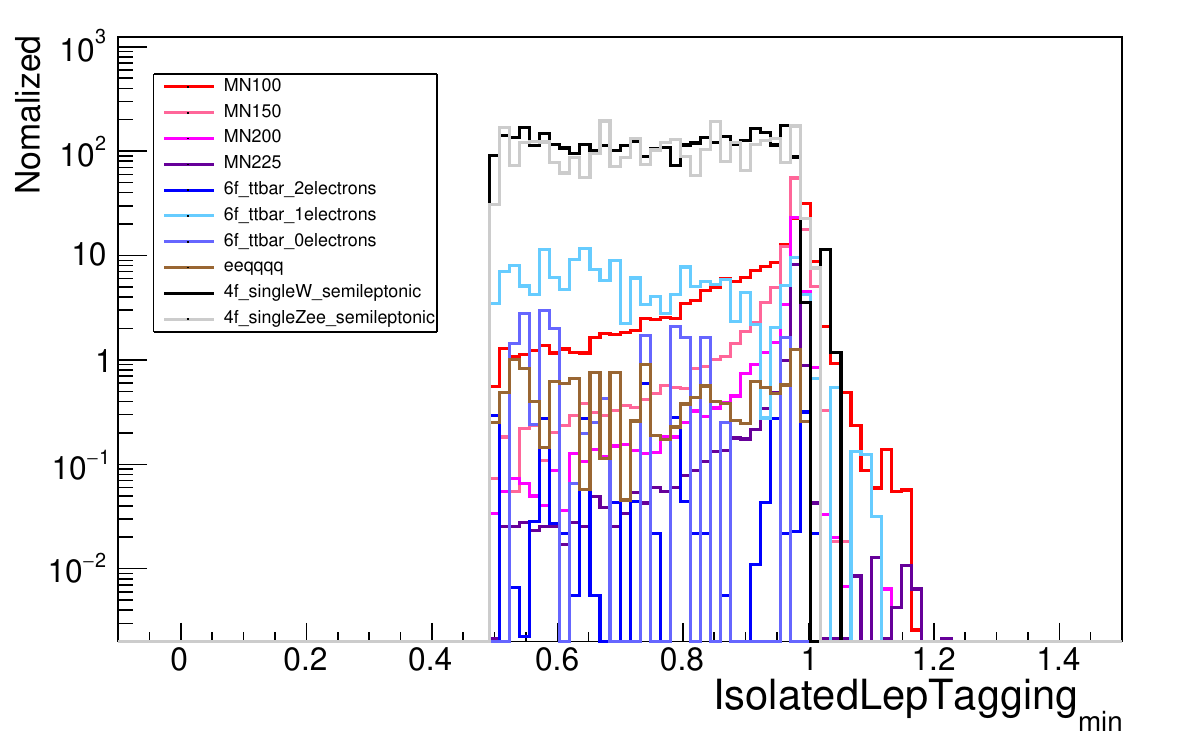} \\
\vspace{-55mm}
\hspace{30mm}{\small    ILD preliminary \hspace{0.3\textwidth}  ILD preliminary }\\
\vspace{55mm}
    \includegraphics[width=0.48\textwidth]{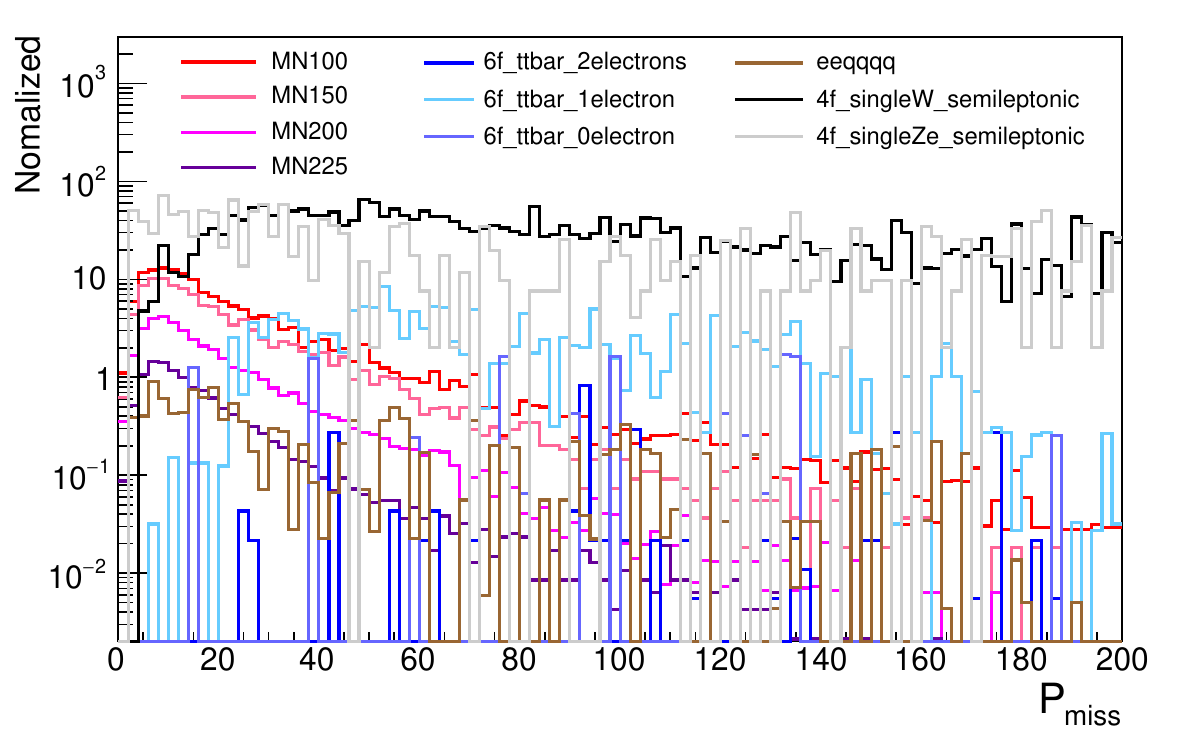} \\
\vspace{-55mm}
{\small  \hspace{0.2\textwidth}  ILD preliminary }\\
\vspace{55mm}
    \caption{Distributions of some cut observables after full simulation and reconstruction.
Product of electron charges; electron energy; electron polar angle; IsolatedTaggingProcesser output; missing momentum.
The signal processes are shown in red/purple/pink. 
Normalized to 1600 $\mathrm{fb^{-1}}$ with \realerpl.}
    \label{fig:cutRL}
 \end{figure}

Distributions of the some of the observables are shown in Fig.~\ref{fig:cutRL}, 
scaled to the integrated luminosity expected at ILC500 ($1600~fb^{-1}$ for each of \realelpr\ \realerpl).

Table~\ref{table:cutflowLR} shows the effect of the cuts in the \realelpr\ and \realerpl\ polarization data. 
Many backgrounds involve the $W$ boson, so more background events remain in the case of \realelpr\ than in \realerpl:
in the case of \realelpr\ 158 SM background events remain, while only 24 remain for \realerpl. 

\begin{table}
  \centering
\begin{tabular}{l | rrrr | rrrrrr}
         &  \multicolumn{4}{c | }{Signal} & \multicolumn{5}{c}{Backgrounds}\\ 
\hline
         &   \multicolumn{4}{c | }{$M_{N}~\mathrm{[GeV]}$} & \multicolumn{2}{c|}{4--fermion} & \multicolumn{4}{c}{6--fermion}        \\ 
\hline
         &  100 & 150 & 200 & 225 &             $e\nu qq$ &              $eeqq$ &   $eeqqqq$     &   $ t\bar{t} (2e) $ & $ t\bar{t} (1e)$ & $ t\bar{t} (0e)$ \\ 
\hline
\hline
         \multicolumn{10}{c}{\realelpr, 1600 $\mathrm{fb^{-1}}$}\\ 
         \hline
 No cut                                        & 559 & 364 & 141 & 46 & 7.77M  & 3.01M & 20k & 19k & 0.30M & 1.15M  \\ 
$\#e_{iso},\mu_{iso},\gamma_{iso}$             & 275 & 190 & 70  & 22 & 90.8k   & 163k & 4.7k  & 9.4k  & 2.2k   & 201   \\
same sign e                                    & 136 & 94  & 34  & 11 & 46.0k   & 3.6k   & 36    & 8     & 430    & 25    \\
electron E                                     & 136 & 94  & 34  & 11 & 41.2k   & 3.3k   & 36    & 8     & 430    & 15    \\
$\cos{\theta_{isoe}}$                          & 126 & 87  & 31  & 10 & 17.5k   & 606    & 12    & 4     & 261    & 15    \\
electron MVA                                   & 79  & 77  & 28  & 9  & 2.6k    & 128    & 2     & 1     & 50     & 0     \\
$y_{12}$                                       & 79  & 77  & 27  & 8  & 501     & 70     & 2     & 0     & 50     & 0     \\
$P_{miss},\cos{\theta_{P_{miss}}}$             & 72  & 72  & 26  & 8  & 118     & 30     & 1     & 0     & 9      & 0     \\ 
\hline
\hline
         \multicolumn{10}{c}{\realerpl, 1600 $\mathrm{fb^{-1}}$}\\ 
\hline
  No cut                                       & 695 & 438 & 161 & 51 & 0.92M & 2.81M & 8.48k  &  8.27k & 0.13M  & 0.50M  \\ 
$\#e_{iso},\mu_{iso},\gamma_{iso}$             & 420 & 343 & 126 & 40 & 7.2k   & 140k & 1.2k  & 3.9k & 870    & 94    \\
same sign e                                    & 164 & 114 & 40  & 12 & 3.7k   & 3.1k   & 14    & 2    & 173    & 12    \\
electron E                                     & 164 & 113 & 40  & 12 & 3.4k   & 2.9k   & 14    & 2    & 173    & 12    \\
$\cos{\theta_{isoe}}$                          & 149 & 104 & 36  & 11 & 1.3k   & 450    & 3     & 1    & 111    & 9     \\
electron MVA                                   & 96  & 93  & 32  & 10 & 198    & 101    & 0     & 0    & 15     & 1     \\
$y_{12}$                                       & 95  & 93  & 32  & 9  & 62     & 32     & 0     & 0    & 15     & 1     \\
$P_{miss},\cos{\theta_{P_{miss}}}$             & 86  & 87  & 30  & 9  & 7      & 15     & 0     & 0    & 2      & 0     \\ 
\hline
\end{tabular}
 \caption{Expected number of events remaining for the 4 benchmark points and SM background, after sequentially applying the cuts described in the text.}
 \label{table:cutflowLR}
\end{table}



In this study, RHN pair production has 4 jets in the final state. After removing isolated $e,~\mu,~\gamma$, particles are forced into 4 jets. 
To reconstruct the RHN mass, we search for the best jet pairing combination.
We assume that the jet pair masses $M_{jj}$ should be consistent with the $W$ mass, and choose the jet pairing which minimizes
\begin{equation}
    F_{1}=(M_{jj1}-M_{w})^{2}+(M_{jj2}-M_{w})^{2}.
    \label{jetpair_W}
\end{equation}

We also assume that the two reconstructed RHN masses $M_{jje1}$, $M_{jje2}$ should be equal, so we then choose the electron--jet--pair combination which minimizes
\begin{equation}
   F_{2}=(M_{jje1}-M_{jje2})^{2}.
    \label{jetpair_RHN}
\end{equation}

\begin{figure}[htb]
\begin{center}
    \includegraphics[width=0.9\textwidth]{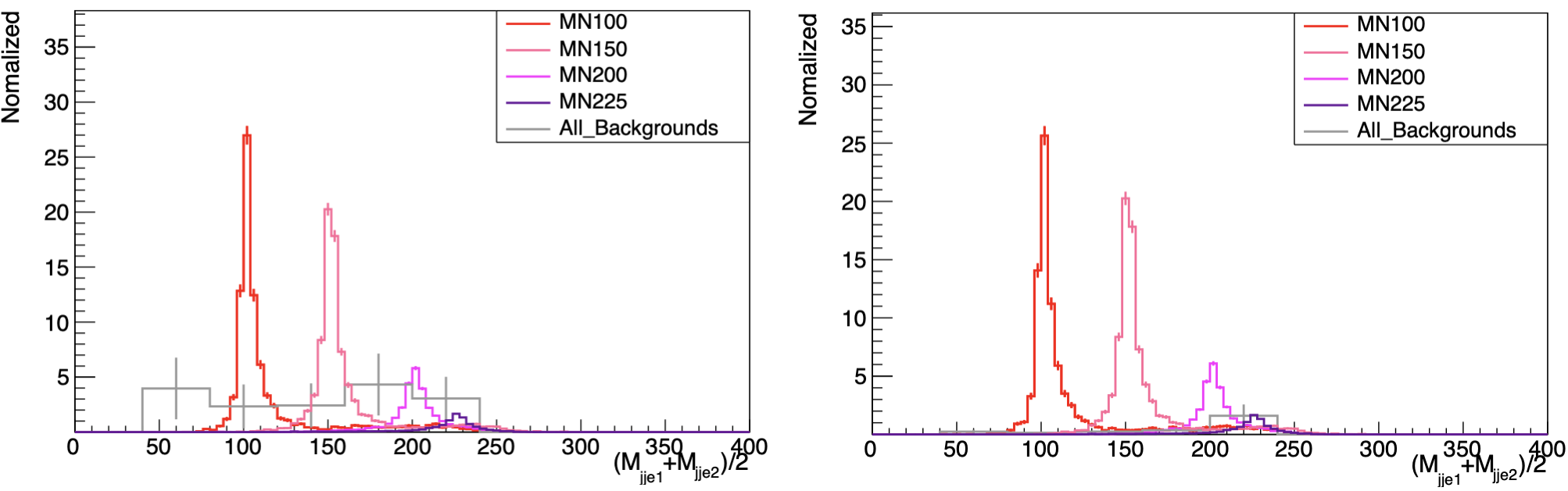} \\
\end{center}
\vspace{-58mm}
\hspace{30mm} {\small ILD preliminary \hspace{45mm} ILD preliminary }\\
\vspace{58mm}
\centering
    \caption{The reconstructed average RHN mass, in signal and background. 
The combined background distribution has been rebinned to reduce the visual effect of fluctuations due to limited MC statistics.)
Left: \realelpr, Right: \realerpl}
    \label{fig:recoRHN}
\end{figure}

Figure~\ref{fig:recoRHN} shows the average of the two reconstructed RHN masses in signal and background, in the two polarisations.
The background is consistent with a flat distribution in the relevant range of this mass observable, while the signals are strongly peaked
around the RHN mass for each benchmark point.

\subsection{Results}

\begin{figure}[htb]
    \centering
    \includegraphics[scale=0.75]{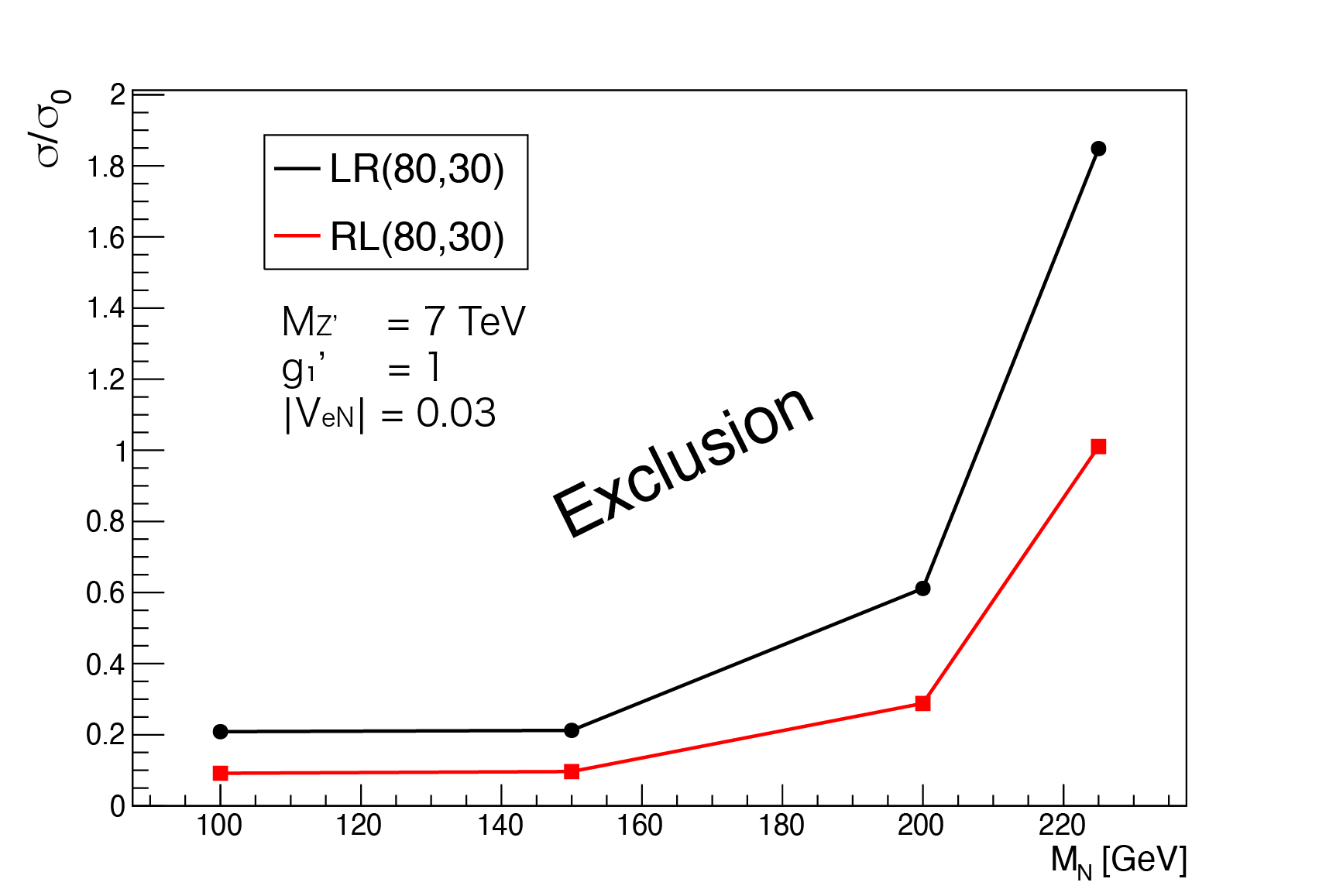}\\
    \vspace{-100mm}
    {\large \hspace{8cm} ILD preliminary}\\
    \vspace{100mm}
    \caption{The obtained 95$\%$ upper limits on the partial cross-section 
$\sigma = \sigma(ee\rightarrow NN)\times (BR(N\rightarrow e^{\pm}W^{\mp}))^{2}$ 
normalised to the benchmark points' cross-sections $\sigma_0$, 
as a function of $M_{N}$. We assume $1600~\mathrm{fb^{-1}}$ of data at ILC-500
in beam polarizations \realelpr\ (black) and \realerpl\ (red).}
    \label{fig:exclusion_sigma}
\end{figure}

In addition to the previous cuts, a mass window from $-10~\mathrm{GeV}$ to $+15~\mathrm{GeV}$ around each true RHN mass was applied. 
The background distribution was assumed to be flat.
%
%
Table~\ref{table:masswindow} 
shows the number of signal and background events remaining after applying the mass window cut.
We use the above results on the number of selected signal and background events $N_{S}, N_{B}$ to calculate the signal significance ($N_{S} / \sqrt{N_{B}+N_{S}}$) 
and expected 95\% confidence upper limits on the partial cross-section. The table also shows the calculated limits, which are presented graphically in Fig.~\ref{fig:exclusion_sigma}.
The cross-section limits are significantly stronger for the \realerpl\ polarization, as expected due to the larger signal cross-section and smaller background.

\begin{table}[htb]
  \centering
\begin{tabular}{ccccc}
\hline
          $M_{N}~\mathrm{[GeV]}$ &   $N_S$          & $N_B$ & $N_{S} / \sqrt{N_{B}+N_{S}}$ & $\sigma^{95} / \sigma_{0}$ \\ 
 \hline
 \hline
\multicolumn{5}{c}{\realelpr} \\
 \hline
100  &  53 & 20.12 & 6.25  & 0.21 \\ 
150  &  52 & 20.12 & 6.18  & 0.21 \\ 
225  &  18 & 20.12 & 2.95  & 0.61 \\ 
250  &  5  & 20.12 & 1.18  & 1.8  \\ 
 \hline
\hline
\multicolumn{5}{c}{\realerpl} \\
 \hline
100  &  66 & 3.24 & 7.98  & 0.092 \\ 
150  &  63 & 3.24 & 7.77  & 0.097 \\ 
225  &  21 & 3.24 & 4.29  & 0.29  \\ 
250  &  6  & 3.24 & 1.99  & 1.0   \\ 
 \hline
\end{tabular}
 \caption{Remaining events after the mass window cuts, signal significance, and 95$\%$ confidence limit exclusion on cross-section $\sigma^{95}$.
Results assume 1600 $\mathrm{fb^{-1}}$ in the \realelpr\ and \realerpl\ polarizations at ILC-500.}
 \label{table:masswindow}
\end{table}

The cross-section limits in two different beam polarization setups will allow the constraining of
the various parameters of the $U(1)_{\rm B-L}$ extension considered in this paper, thanks to the
different dependencies of the $s$- and $t$-channel diagrams on the model parameters 
${M_{N}}$, ${M_{Z^{\prime}}}$, $g_{B-L}^{\prime}$, $|{V}_{eN}|$. We plan to investigate this further
during the Snowmass study.

\section{Summary}

An experimental study of Majorana RHN pair production in the context of a $U(1)_{\rm B-L}$ extension of the SM was performed using full simulation of ILD at the 500~GeV stage of ILC.
This process proceeds via two diagrams at tree level: $s$-channel $Z^\prime$ exchange or $t$-channel $W$ exchange. The relative size contributions
depend both on the parameters of the underlying model and on the ILC's beam polarization.
For the benchmarks studied in this paper, the signal cross-section is larger with predominantly right-handed electrons / left-handed positrons,
in which case the $s$-channel dominates. The signal cross-section in the other polarization is suppressed by strong negative interference between the diagrams.
The SM backgrounds, which often involve a $W$ boson, are suppressed when right-handed electrons / left-handed positrons are used.

The decay of both RHNs to electron+$W$ provides a unique and almost background-free signature when the two electrons have the same electric charge.
Standard Model backgrounds in the same-sign electron, low missing momentum, and 4--jet final state are extremely small.
RHN masses in the range 100--225~GeV were studied in the context of a minimal B-L model. 
We considered the dominant 2- and 4-fermion backgrounds, and find that after the event selection only 3 (20) background events
remain in $1600 fb^{-1}$ of integrated luminosity in the \realerpl\ (\realelpr) polarization, allowing the
setting of cross-section upper limits up to 10 times smaller than those of the chosen benchmarks.

\section*{Acknowledgments}

We would like to thank the LCC generator working group and the ILD software working group for providing the simulation and reconstruction tools and producing the Monte Carlo samples used in this study.
This work has benefited from computing services provided by the ILC Virtual Organization, supported by the national resource providers of the EGI Federation and the Open Science GRID.
This work was supported in part by JSPS KAKENHI Grant Number JP21H01077 (D.J.), and United States Department of Energy Grant No. DE-SE0012447 (N.O.).

\bibliographystyle{unsrt}


\begin{thebibliography}{10}

\bibitem{Patrignani:2016xqp}
C.~Patrignani et~al.
\newblock {Review of Particle Physics}.
\newblock {\em Chin. Phys.}, C40(10):100001, 2016.

\bibitem{Minkowski:1977sc}
Peter Minkowski.
\newblock {$\mu \to e\gamma$ at a Rate of One Out of $10^{9}$ Muon Decays?}
\newblock {\em Phys. Lett.}, 67B:421--428, 1977.

\bibitem{Mohapatra:1979ia}
Rabindra~N. Mohapatra and Goran Senjanovic.
\newblock {Neutrino Mass and Spontaneous Parity Nonconservation}.
\newblock {\em Phys. Rev. Lett.}, 44:912, 1980.
\newblock [,231(1979)].

\bibitem{Schechter:1980gr}
J.~Schechter and J.~W.~F. Valle.
\newblock {Neutrino Masses in SU(2) x U(1) Theories}.
\newblock {\em Phys. Rev.}, D22:2227, 1980.

\bibitem{Yanagida:1979as}
Tsutomu Yanagida.
\newblock {Horizontal gauge symmetry and masses of neutrinos}.
\newblock {\em Conf. Proc.}, C7902131:95--99, 1979.

\bibitem{GellMann:1980vs}
Murray Gell-Mann, Pierre Ramond, and Richard Slansky.
\newblock {Complex Spinors and Unified Theories}.
\newblock {\em Conf. Proc.}, C790927:315--321, 1979.

\bibitem{Glashow:1979nm}
S.~L. Glashow.
\newblock {The Future of Elementary Particle Physics}.
\newblock {\em NATO Sci. Ser. B}, 61:687, 1980.

\bibitem{Kang:2015uoc}
Zhaofeng Kang, P.~Ko, and Jinmian Li.
\newblock {New Avenues to Heavy Right-handed Neutrinos with Pair Production at
  Hadronic Colliders}.
\newblock {\em Phys. Rev. D}, 93(7):075037, 2016.

\bibitem{Cox:2017eme}
Peter Cox, Chengcheng Han, and Tsutomu~T. Yanagida.
\newblock {LHC Search for Right-handed Neutrinos in $Z^\prime$ Models}.
\newblock {\em JHEP}, 01:037, 2018.

\bibitem{Accomando:2017qcs}
Elena Accomando, Luigi Delle~Rose, Stefano Moretti, Emmanuel Olaiya, and
  Claire~H. Shepherd-Themistocleous.
\newblock {Extra Higgs boson and Z' as portals to signatures of heavy neutrinos
  at the LHC}.
\newblock {\em JHEP}, 02:109, 2018.

\bibitem{Das:2017flq}
Arindam Das, Nobuchika Okada, and Digesh Raut.
\newblock {Enhanced pair production of heavy Majorana neutrinos at the LHC}.
\newblock {\em Phys. Rev. D}, 97(11):115023, 2018.

\bibitem{Das:2017deo}
Arindam Das, Nobuchika Okada, and Digesh Raut.
\newblock {Heavy Majorana neutrino pair productions at the LHC in minimal U(1)
  extended Standard Model}.
\newblock {\em Eur. Phys. J. C}, 78(9):696, 2018.

\bibitem{Jana:2018rdf}
Sudip Jana, Nobuchika Okada, and Digesh Raut.
\newblock {Displaced vertex signature of type-I seesaw model}.
\newblock {\em Phys. Rev. D}, 98(3):035023, 2018.

\bibitem{Das:2019fee}
Arindam Das, P.~S.~Bhupal Dev, and Nobuchika Okada.
\newblock {Long-lived TeV-scale right-handed neutrino production at the LHC in
  gauged $U(1)_X$ model}.
\newblock {\em Phys. Lett. B}, 799:135052, 2019.

\bibitem{Chiang:2019ajm}
Cheng-Wei Chiang, Giovanna Cottin, Arindam Das, and Sanjoy Mandal.
\newblock {Displaced heavy neutrinos from $Z'$ decays at the LHC}.
\newblock {\em JHEP}, 12:070, 2019.

\bibitem{Liu:2022kid}
\newblock W.~Liu, S.~Kulkarni and F.~F.~Deppisch,
\newblock {Heavy Neutrinos at the FCC-hh in the $U(1)_{B-L}$ Model}.
\newblock [arXiv:2202.07310 [hep-ph]].

\bibitem{Das:2018tbd}
Arindam Das, Nobuchika Okada, Satomi Okada, and Digesh Raut.
\newblock {Probing the seesaw mechanism at the 250 GeV ILC}.
\newblock {\em Phys. Lett. B}, 797:134849, 2019.

\bibitem{ATLAS:2019erb}
\newblock G.~Aad \textit{et al.} [ATLAS],
\newblock {Search for high-mass dilepton resonances using 139 fb$^{-1}$ of $pp$ collision data collected at $\sqrt{s}=$13 TeV with the ATLAS detector}.
\newblock Phys. Lett. B \textbf{796}, 68-87 (2019)
\newblock [arXiv:1903.06248 [hep-ex]].

\bibitem{CMS:2021ctt}
Albert~M Sirunyan et~al.
\newblock {Search for resonant and nonresonant new phenomena in high-mass
  dilepton final states at $ \sqrt{s} $ = 13 TeV}.
\newblock {\em JHEP}, 07:208, 2021.

\bibitem{CERN-LHCC-2017-018}
{Technical Design Report for the Phase-II Upgrade of the ATLAS LAr
  Calorimeter}.
\newblock Technical Report CERN-LHCC-2017-018. ATLAS-TDR-027, CERN, Geneva, Sep
  2017.

\bibitem{ATLAS:2019bov}
\newblock  [ATLAS],
\newblock {Search for New Phenomena in Dijet Events using 139 fb$^{-1}$ of $pp$ collisions at $\sqrt{s}$ = 13TeV collected with the ATLAS Detector}.
\newblock ATLAS-CONF-2019-007.

\bibitem{Sirunyan:2018xlo}
Albert~M Sirunyan et~al.
\newblock {Search for narrow and broad dijet resonances in proton-proton
  collisions at $ \sqrt{s}=13 $ TeV and constraints on dark matter mediators
  and other new particles}.
\newblock {\em JHEP}, 08:130, 2018.

\bibitem{Electroweak:2003ram}
\newblock [LEP, ALEPH, DELPHI, L3, OPAL, LEP Electroweak Working Group, SLD Electroweak Group and SLD Heavy Flavor Group],
\newblock {A Combination of preliminary electroweak measurements and constraints on the standard model}.
\newblock [arXiv:hep-ex/0312023 [hep-ex]].

\bibitem{LCCPhysicsWorkingGroup:2019fvj}
\newblock K.~Fujii \textit{et al.} [LCC Physics Working Group],
\newblock {Tests of the Standard Model at the International Linear Collider}.
\newblock [arXiv:1908.11299 [hep-ex]].

\bibitem{Carena:2004xs}
Marcela Carena, Alejandro Daleo, Bogdan~A. Dobrescu, and Timothy M.~P. Tait.
\newblock {$Z^\prime$ gauge bosons at the Tevatron}.
\newblock {\em Phys. Rev.}, D70:093009, 2004.

\bibitem{Okada:2016tci}
Nobuchika Okada and Satomi Okada.
\newblock {$Z^\prime$-portal right-handed neutrino dark matter in the minimal
  U(1)$_X$ extended Standard Model}.
\newblock {\em Phys. Rev. D}, 95(3):035025, 2017.

\bibitem{Das:2021esm}
A.~Das, P.~S.~B.~Dev, Y.~Hosotani and S.~Mandal,
{Probing the minimal $U(1)_X$ model at future electron-positron colliders via the fermion pair-production channel}.
[arXiv:2104.10902 [hep-ph]].


\bibitem{Schael:2013ita}
S.~Schael et~al.
\newblock {Electroweak Measurements in Electron-Positron Collisions at
  W-Boson-Pair Energies at LEP}.
\newblock {\em Phys. Rept.}, 532:119--244, 2013.

\bibitem{Alwall:2011uj}
Johan Alwall, Michel Herquet, Fabio Maltoni, Olivier Mattelaer, and Tim
  Stelzer.
\newblock {MadGraph 5 : Going Beyond}.
\newblock {\em JHEP}, 06:128, 2011.

\bibitem{Alwall:2014hca}
J.~Alwall, R.~Frederix, S.~Frixione, V.~Hirschi, F.~Maltoni, O.~Mattelaer,
  H.~S. Shao, T.~Stelzer, P.~Torrielli, and M.~Zaro.
\newblock {The automated computation of tree-level and next-to-leading order
  differential cross sections, and their matching to parton shower
  simulations}.
\newblock {\em JHEP}, 07:079, 2014.

\bibitem{Behnke:2013xla}
T.~Behnke, J.~E.~Brau, B.~Foster, J.~Fuster, M.~Harrison, J.~M.~Paterson, M.~Peskin, M.~Stanitzki, N.~Walker and H.~Yamamoto,
{The International Linear Collider Technical Design Report - Volume 1: Executive Summary}.
[arXiv:1306.6327 [physics.acc-ph]].

\bibitem{ILDConceptGroup:2020sfq}
H.~Abramowicz \textit{et al.} [ILD Concept Group],
{International Large Detector: Interim Design Report}.
[arXiv:2003.01116 [physics.ins-det]].

\bibitem{2011_Whizard}
Wolfgang Kilian, Thorsten Ohl, and Jürgen Reuter.
\newblock {WHIZARD—simulating multi-particle processes at LHC and ILC}.
\newblock {\em The European Physical Journal C}, 71(9), Sep 2011.

\bibitem{Das:2022rbl}
A.~Das, S.~Mandal, T.~Nomura and S.~Shil,
\newblock {Heavy Majorana neutrino pair production from $Z^\prime$ at hadron and lepton colliders}.
\newblock [arXiv:2202.13358 [hep-ph]].

\bibitem{Schulte:1999tx}
D.~Schulte,
\newblock {Beam-beam simulations with GUINEA-PIG}.
\newblock CERN-PS-99-014-LP, CERN-CLIC-NOTE-387.

\bibitem{adolphsen2013international}
C.~Adolphsen, M.~Barone, B.~Barish, K.~Buesser, P.~Burrows, J.~Carwardine, J.~Clark, H.~Mainaud Durand, G.~Dugan and E.~Elsen, \textit{et al.}
{The International Linear Collider Technical Design Report - Volume 3.II: Accelerator Baseline Design}.
[arXiv:1306.6328 [physics.acc-ph]].

\bibitem{Sjostrand:2006za}
T.~Sjostrand, S.~Mrenna and P.~Skands,
\newblock {PYTHIA 6.4 Physics and Manual}
\newblock [arXiv:hep-ph/0603175].

\bibitem{sailer2017dd4hep}
Andre Sailer, Markus Frank, Frank Gaede, Daniel Hynds, Shaojun Lu, Nikiforos
  Nikiforou, Marko Petric, Rosa Simoniello, Georgios Voutsinas, et~al.
\newblock {DD4Hep based event reconstruction}.
\newblock In {\em Journal of Physics: Conference Series}, volume 898, page
  042017. IOP Publishing, 2017.


\bibitem{GEANT4:2002zbu}
S.~Agostinelli \textit{et al.} [GEANT4],
{GEANT4--a simulation toolkit}.
Nucl. Instrum. Meth. A \textbf{506}, 250-303 (2003)

\bibitem{Gaede:2006pj}
F.~Gaede,
{Marlin and LCCD: Software tools for the ILC,}
Nucl. Instrum. Meth. A \textbf{559}, 177-180 (2006)

\bibitem{Thomson:2009rp}
M.~A.~Thomson,
{Particle Flow Calorimetry and the PandoraPFA Algorithm.}
Nucl. Instrum. Meth. A \textbf{611}, 25-40 (2009)

\bibitem{Catani1991432}
S.~Catani, Yu.L. Dokshitzer, M.~Olsson, G.~Turnock, and B.R. Webber,
\newblock {New clustering algorithm for multijet cross sections in e+e-
  annihilation}.
\newblock {\em Physics Letters B}, 269(3-4):432--438, 1991.

\end{thebibliography}

\end{document}